\documentclass[12pt]{amsart}

\usepackage{ljm-auth}

\usepackage{amssymb}
\usepackage{amsmath}
\usepackage{amsthm}
\usepackage{textcomp}
\usepackage{array}
\usepackage[utf8]{inputenc}
\usepackage[english]{babel}

\usepackage{graphicx}

\usepackage{hyperref}

\author{K.\,Khadiev}
\crauthor{K.\,Khadiev} \tit{Width Hierarchy for $k$-OBDD of Small
Width}

\newtheorem{theorem}{Theorem}
\newtheorem{lemma}{Lemma}
\newcommand{\Endproof}{\hfill$\Box$\\}
\newcommand{\Beginproof}{{\em Proof.\quad}}

\begin{document}

\maketit
\address{Kazan Federal University, Russia}

\email{kamilhadi@gmail.com}

\abstract{In this paper was explored well known model $k$-OBDD.
There are proven width based hierarchy of classes of boolean
functions which computed by $k$-OBDD. The proof of hierarchy is
based on sufficient condition of Boolean function's non
representation as $k$-OBDD and complexity properties of Boolean
function SAF. This function is modification of known Pointer Jumping
(PJ) and Indirect Storage Access (ISA) functions.}

\notes{0}{
\subclass{}%
\keywords{Branching programs, Binary decision diagrams, OBDD, k-OBDD,  complexity classes}%
\thank{Partially supported by Russian Foundation for Basic Research,
Grant 14-07-00557. The work is performed according to the Russian Government Program of Competitive Growth of Kazan Federal University}
}

\section{Preliminaries}

The  $k$-OBDD and OBDD models  are well known models of branching
programs. Good source for a different models of branching programs
is the book by Ingo Wegener  \cite{we00}.

The branching program $P$ over a set $X$ of $n$ Boolean variables is
a directed acyclic graph with a source node and sink nodes. Sink
nodes are labeled by $1$ (Accept) or $0$ (Reject). Each inner node
$v$ is associated with a variable $x\in X$ and has two outgoing
edges labeled $x=0$   and $x=1$ respectively. An input $\nu\in
\{0,1\}^n$ determines  a computation (consistent) path of  from the
source node of $P$ to a one of the sink nodes of $P$. We denote
$P(\nu)$ the label of sink finally reached by $P$ on the input
$\nu$. The input $\nu$  is accepted or rejected if $P(\nu)=1$ or
$P(\nu)=0$ respectively.

Program $P$ computes (presents) Boolean function $f(X)$
($f:\{0,1\}^n \rightarrow \{0,1\}$) if $f(\nu)=P(\nu)$ for all $\nu
\in \{0,1\}^n$.

A branching program is {\em leveled} if the nodes can be partitioned into levels $V_1, \ldots,
V_\ell$ and a level $V_{\ell+1}$ such that the nodes in $V_{\ell+1}$
are the sink nodes, nodes in each level $V_j$ with $j \le \ell$ have
outgoing edges only to nodes in the next level $V_{j+1}$.

The {\em width} $w(P)$ of leveled branching program $P$ is the
maximum of number of nodes in  levels of $P$: $  w(P)=\max_{1\le
j\le \ell}|V_j|.$

A leveled branching program is called {\em oblivious} if all inner
nodes of one level are labeled by the same variable.  A branching
program is called {\em read once} if each variable is tested on each
path only once.

The oblivious leveled read once branching program is also called
Ordinary Binary Decision Diagram (OBDD).

A branching program $P$ is called $k$-OBDD with order $\theta(P)$ if
it consists of $k$ layers and each $i$-th layer is OBDD with the
same order $\theta(P)$. In nondeterministic case it is denoted
$k$-NOBDD.

The {\em size} $s(P)$ of branching program $P$ is a number of nodes
of program $P$. Note, that for $k$-OBDD and $k$-NOBDD following is
right: $ s(P)<w(P) \cdot n \cdot k $.

There are many paper which explore width and size as measure of complexity of classes.
 Most of them investigate exponential difference between models of Branching Program. Models with less restrictions than $k$-OBDD
  like  non-deterministic, probabilistic and others also were explored, for example
  in papers \cite{brs93, AGKMP05, A97, ak98, bssw96, hs2000, hs2003, sau2000, tha98}.
More precise width hierarchy is presented in the paper.

We denote $\mathsf{k-OBDD_{w}}$ is the sets of Boolean functions
that have representation as $k$-OBDD  of width $w$. We denote
$\mathsf{k-OBDD_{POLY}}$ and $\mathsf{k-OBDD_{EXP}}$ is the sets of
Boolean functions that have representation as $k$-OBDD of polynomial
and exponential width respectively. In  \cite{bssw96} was shown that
$\mathsf{k-OBDD_{POLY}}\subsetneq\mathsf{k-OBDD_{EXP}}$. Result in
this paper is following.

\begin{theorem}\label{h-kobdd}
For integer $k=k(n),w=w(n)$  such that $2kw(2w + \lceil \log k
\rceil + \lceil \log 2w \rceil)<n, k\geq 2, w\geq 64$ we have
$\mathsf{k-OBDD_{\lfloor w/16
\rfloor-3}}\subsetneq\mathsf{k-OBDD_{w}}$.
\end{theorem}

Analogosly hierarchies was considered for OBDD in paper
\cite{agky14} and for two way non-uniform automata in cite{ky14}.
This kind of automata can be considered like special type of
branching programs.

Proof of this Theorem is presented in following section. It based on lower bound which presented in \cite{ak13}.

\section{Proof of Theorem \ref{h-kobdd}}

We start with needed  definitions and notations.

 Let $\pi=(\{x_{j_1},\dots, x_{j_u}\}, \{x_{i_1},\dots, x_{i_v}\})=(X_A,X_B)$
 be a partition of the set $X$ into two parts $X_A$ and $X_B=X\backslash X_A$. Below we will use equivalent notations $f(X)$ and $f(X_A, X_B)$.

Let  $f|_\rho$ be subfunction of $f$, where  $\rho$ is mapping $\rho:X_A \to \{0,1\}^{|X_A|}$.
Function $f|_\rho$ is obtained from $f$ by applying $\rho$. We denote $N^\pi(f)$ to be amount of different subfunctions with respect to partition $\pi$.

Let $\Theta(n)$ be the set of all permutations of $\{1,\dots,n\}$.
We say, that  partition $\pi$ agrees with permutation
$\theta=(j_1,\dots, j_n)\in \Theta(n)$, if for some $u$, $1<u<n$ the
following is right: $\pi=(\{x_{j_1},\dots,
x_{j_u}\},\{x_{j_{u+1}},\dots, x_{j_n}\})$. We denote $\Pi(\theta)$
a set of all partitions which agrees with $\theta$.

Let $ N^\theta(f)=   \max_{\pi\in \Pi(\theta)} N^\pi(f), \qquad
N(f)=\min_{\theta\in \Theta(n)}N^\theta(f). $ Proof of Theorem
\ref{h-kobdd} based on following Lemmas and complexity properties of
Boolean {\em Shuffled Address Function} $SAF_{k,w}(X)$.

Let us  define Boolean function $SAF_{k,w}(X):\{0,1\}^n\to \{0,1\}$ for integer $k=k(n)$ and  $w=w(n)$ such that
\begin{equation}
 2kw(2w + \lceil \log k \rceil + \lceil \log 2w \rceil)<n.\label{kw}
\end{equation}
We divide input variables to $2kw$ blocks. There are $\lceil
n/(2kw)\rceil =a$ variables in each block.  After that we divide
each block to
 {\em address} and {\em value} variables. First  $\lceil\log k\rceil + \lceil\log 2w\rceil$ variables of block are {\em address}
 and other $a-\lceil\log k\rceil + \lceil\log 2w\rceil=b$ variables of block are {\em value}.

We call $x^{p}_{0},\dots,x^{p}_{b-1}$ {\em value} variables of $p$-th block and  $y^{p}_{0},\dots,y^{p}_{\lceil\log k\rceil + \lceil\log 2w\rceil}$ are {\em address} variables, for $p\in\{0,\dots,2kw-1\}$.

Boolean function $SAF_{k,w}(X)$ is iterative  process  based on definition of following six functions:

Function $AdrK:\{0,1\}^n\times\{0,\dots,2kw-1\}\to \{0,\dots,k-1\}$
obtains firsts part of block's address. This block will be used only
in step of iteration which number is computed using this function:

\[AdrK(X,p)=\sum_{j=0}^{\lceil\log k\rceil-1}y^{p}_{j}\cdot 2^{j} (mod\textrm{
}k).\]

Function $AdrW:\{0,1\}^n\times\{0,\dots,2kw-1\}\to \{0,\dots,2w-1\}$ obtains second part of block's address. It is the address of block within one step of iteration:

\[AdrW(X,p)=\sum_{j=0}^{\lceil\log 2w\rceil-1}y^{p}_{j+\lceil\log k\rceil}\cdot 2^{j} (mod\textrm{
}2w).\]

Function $Ind:\{0,1\}^n\times\{0,\dots,2w-1\}\times\{0,\dots,k-1\}\to \{0,\dots,2kw-1\}$ obtains number of block by number of step and address within this step of iteration:

\begin{displaymath}
Ind(X,i,t) = \left\{ \begin{array}{ll}
p, & \textrm{where $p$ is minimal number of block such that}\\
& \textrm{$AdrK(X,p)=t$ and $AdrW(X,p)=i$}, \\
-1, & \textrm{if there are no such $p$}.
\end{array} \right.
\end{displaymath}

Function $Val:\{0,1\}^n\times\{0,\dots,2w-1\}\times\{1,\dots,k\}\to \{-1,\dots,w-1\}$ obtains value of block which have address $i$ within $t$-th step of iteration:

\begin{displaymath}
Val(X,i,t) = \left\{ \begin{array}{ll}
\sum_{j=0}^{b-1}x^{p}_{j} (mod\textrm{ }w), & \textrm{where }p=Ind(X,i,t)\textrm{, for $p\geq 0$}, \\
-1, & \textrm{if }Ind(X,i,t)<0.
\end{array} \right.
\end{displaymath}

Two functions $Step_1$ and $Step_2$ obtain value of $t$-th step of iteration. Function $Step_1:\{0,1\}^n\times\{0,\dots,k-1\}\to \{-1,w\dots,2w-1\}$ obtains base for value of step of iteration:

\begin{displaymath}
Step_1(X,t) = \left\{ \begin{array}{ll}
-1, & \textrm{if }  Step_2(X,t-1)=-1, \\
0, & \textrm{if }  t=-1,\\
Val(X,Step_2(X,t-1),t) + w, & \textrm{otherwise}.
\end{array} \right.
\end{displaymath}

Function $Step_2:\{0,1\}^n\times\{0,\dots,k-1\}\to \{-1,\dots,w-1\}$ obtain value of $t$-th step of iteration:

\begin{displaymath}
Step_2(X,t) = \left\{ \begin{array}{ll}
-1, & \textrm{if }  Step_1(X,t)=-1, \\
0, & \textrm{if }  t=-1\\
Val(X,Step_1(X,t),t), & \textrm{otherwise}.
\end{array} \right.
\end{displaymath}

Note that address of current block is computed on previous step.

Result of Boolean function $SAF_{k,w}(X)$ is computed by following way:

\begin{displaymath}
SAF_{k,w}(X) = \left\{ \begin{array}{ll}
0, & \textrm{if }  Step_2(X,k-1)\leq 0, \\
1, & \textrm{otherwise}.
\end{array} \right.
\end{displaymath}

Let us discuss complexity properties of this function in Lemma \ref{fkw1} and Lemma \ref{fkw2}. Proof of Lemma \ref{fkw1} uses  following technical Lemmas \ref{good-set} and \ref{good-input}.

\begin{lemma}\label{good-set}
Let integer $k=k(n)$ and $w=w(n)$ are such that inequality (\ref{kw}) holds. Let partition $\pi=(X_A,X_B)$ is such that $X_A$ contains at least $w$ {\em value} variables from exactly $kw$ blocks. Then $X_B$ contains at least $w$ {\em value} variables from exactly $kw$ blocks.
\end{lemma}
\Beginproof
Let $I_A=\{i:$  $X_A$ contains at least $w$ {\em value} variables from $i$-th block$\}$.
 And let $i'\not\in I_A$ then $X_A$ contains at most $w-1$ {\em value} variables from $i'$-th block.
  Hence $X_B$ contains at least $b-(w-1)$ {\em value} variables from $i'$-th block. By (\ref{kw}) we have:

\[b-(w-1)=(n/(2kw)-(\lceil \log k \rceil + \lceil \log 2w \rceil)-(w-1)>\]\[>(2w + \lceil \log k \rceil +
\lceil \log 2w \rceil)-(\lceil \log k \rceil + \lceil \log 2w \rceil)-(w-1)=2w-(w-1)=w+1.\]

Let set $I=\{0,\dots,2kw-1\}$ is numbers of all blocks and $i'\in
I\backslash I_A$. Note that $|I\backslash I_A|=2kw- kw =kw$.
\Endproof

Let us choose any order $\theta\in \Theta(n)$. And we choose partition $\pi=(X_A,X_B)\in \Pi(\theta)$  such that $X_A$ contains at least $w$ {\em value} variables from exactly $kw$ blocks. Let $I_A=\{i:$  $X_A$ contains at least $w$ {\em value} variables from $i$-th block$\}$ and $I_B=\{0,\dots,2kw-1\}\backslash I_A$. By Lemma \ref{good-set} we have $|I_B|=kw$.

For input $\nu$ we have partition $(\sigma,\gamma)$ with respect to
$\pi$. We define sets $\Sigma\subset\{{0,1\}^{|X_A|}}$ and
$\Gamma\subset\{{0,1\}^{|X_B|}}$ for input with respect to $\pi$,
that satisfies the following conditions: for
$\sigma,\sigma'\in\Sigma$, $\gamma\in\Gamma$ and
$\nu=(\sigma,\gamma)$, $\nu'=(\sigma',\gamma)$ we have

\begin{itemize}
\item for any $r\in\{0,\dots,k-1\}$ and $z\in\{0,\dots, w-1\}$ it is true that $Ind(\nu,z,r)\in I_A$;
\item for any $r\in\{0,\dots,k-1\}$ and $z\in\{w,\dots, 2w-1\}$ it is true that $Ind(\nu,z,r)\in I_B$;
\item there are $r\in\{1,\dots,k-1\}$, $z\in\{0,\dots, w-3\}$, such that $Val(\nu',z,r)\neq
Val(\nu,z,r)$;
\item value of $x^{p}_{j}$ is $0$, for any $p\in I_B$ and $x^{p}_{j}\in X_A$;
\item value of $x^{p}_{j}$ is $0$, for any $p\in I_A$ and $x^{p}_{j}\in X_B$;
\item following statement is right:
\begin{equation}\label{nu1}
Val(\nu,w-2,t)=2w-2, Val(\nu',w-1,t)=2w-1, \textrm{ for }0\leq t\leq
k-1;
\end{equation}
\begin{equation}\label{nu2}
 Val(\nu,2w-2,t)=w-2,Val(\nu,2w-1,t)=w-1  \textrm{ for }0\leq t\leq
 k-2;
\end{equation}
\item for $p=Ind(\nu,2w-1,k-1)$ and $p'=Ind(\nu,2w-2,k-1)$ following statement is right:
\begin{equation}\label{nu3}
Val(\nu, 2w-1,k-1)=0\quad\quad Val(\nu, 2w-2,k-1)=1.
\end{equation}
\end{itemize}

Let us show needed property of this sets.

\begin{lemma}\label{good-input}
Sets $\Sigma$ and $\Gamma$ such that for any sequence
$v=(v_0,\dots,v_{ 2(k-1)(w-2)-1})$, for $v_i\in\{0,\dots, w-1\}$,
there are $\sigma\in \Sigma$ and $\gamma\in\Gamma$ such that: for
each $i\in \{0,\dots, (k-1)(w-2)-1\}$ there are
$r_i\in\{1,\dots,k-1\}$ and $z_i\in\{0,\dots, w-3\}$ such that
$Val(\nu,z_i,r_i)=a_i$, and for each $i\in \{(k-1)(w-2),\dots,
2(k-1)(w-2)-1\}$ there are $r_i\in\{1,\dots,k-1\}$ and
$z_i\in\{w,\dots, 2w-3\}$ such that $Val(\nu,z_i,r_i)=a_i$.
\end{lemma}
\Beginproof
Let $p_i\in I_A$, such that $p_i=Ind(\nu,z_i,r_i)$, for $i\in \{0,\dots, (k-1)(w-2)-1\}$.
Let us remind that value of $x^{p_i}_{j}$ is $0$ for any $x^{p_i}_{j}\in X_B$.
Hence value of $Val(\nu,z_i,r_i)$ depends only on variables from $X_A$. At least $w$ {\em value} variables of $p_i$-th block belong to $X_A$.
Hence we  can choose input $\sigma$ with $a_i$ $1$'s in {\em value} variables of $p_i$-th block which belongs to $X_A$.

Input $\gamma\in\Gamma$ and $i\in \{(k-1)(w-2),\dots, 2(k-1)(w-2)-1\}$ we can proof by the same way.
\Endproof

\begin{lemma}\label{fkw1}
For integer $k=k(n)$, $w=w(n)$ and Boolean function $SAF_{k,w}$,
such that inequality (\ref{kw}) holds, the following statement is
right: $N(SAF_{k,w})\geq w^{(k-1)(w-2)}$.
\end{lemma}

\Beginproof Let us choose any order $\theta\in \Theta(n)$. And we
choose partition $\pi=(X_A,X_B)\in \Pi(\theta)$  such that $X_A$
contains at least $w$ {\em value} variables from exactly $kw$
blocks. Let us consider two different inputs $\sigma,\sigma'\in
\Sigma$ and corresponding mappings $\tau$ and $\tau'$. Let us show
that subfunctions $SAF_{k,w}|_\tau$ and $SAF_{k,w}|_{\tau'}$ are
different. Let $r\in\{1,\dots,k-2\}$ and $z\in\{0,\dots, w-3\}$ are
such that $s'=Val(\nu',z,r)\neq Val(\nu,z,r)=s$. Let us choose
$\gamma\in \Gamma$ such that $Val(\nu,s+w,r)=w-1,$
$Val(\nu',s'+w,r)=w-2$ and  $Val(\nu,i,r-1)=Val(\nu',i,r-1)=z$,
where $i\in \{w,\dots,2w-1\}$.

It means $Step_2(\nu,r-1)=Step_2(\nu',r-1)=z$ and $Step_2(\nu,r)=w-1, Step_2(\nu',r)=w-2$.
 Also conditions (\ref{nu1}), (\ref{nu2}) mean that $Step_2(\nu,t)=w-1, Step_2(\nu',t)=w-2$, for $r< t\leq k$. Hence $Step_1(\nu,k-1)= 2w-2, Step_1(\nu',k-1)= 2w-1$
 and by (\ref{nu3}) we have $SAF_{k,w}(\nu)\neq SAF_{k,w}(\nu')$.

Let $r=k-1$, $z\in\{0,\dots, w-3\}$ such that $s'=Val(\nu',z,r)\neq
Val(\nu,z,r)=s$. Let us choose $\gamma\in \Gamma$ such that
$Val(\nu,s+w,r)=1$ ,$Val(\nu',s'+w,r)=0$. Therefore
$SAF_{k,w}|_\tau(\gamma)\neq SAF_{k,w}|_{\tau'}(\gamma)$ also
$SAF_{k,w}|_\tau\neq SAF_{k,w}|_{\tau'}$.

Let us compute $|\Sigma|$. For $\sigma\in\Sigma$ by Lemma
\ref{good-input} we can get each value of $Val(\nu,i,t)$ for $0\leq
i \leq w-3$ and $1\leq t \leq k-1$. It means $|\Sigma|\geq
w^{(k-1)(w-2)}$. Therefore $N^{\pi}(SAF_{k,w})\geq w^{(k-1)(w-2)}$
and by definition of $N(SAF_{k,w})$ we have $N(SAF_{k,w})\geq
w^{(k-1)(w-2)}$.
\Endproof

\begin{lemma}\label{fkw2}
There is $2k$-OBDD $P$ of width $3w+1$ which computes $SAF_{k,w}$
\end{lemma}
\Beginproof
Let us construct $P$. Let us use natural order $(1,\dots,n)$ and in
each $(2t-1)$-th layer $P$ computes $Step_1(X,t-1)$ and in each $(2t)$-th layer it
computes $Step_2(X,t-1)$. Let us consider computation on input $\nu\in\{0,1\}^n$.

Let us consider layer $2t-1$. The first level contains $w$ nodes for store each value of function $Step_2(\nu,t-2)$. For $i$-th  node
of first level program $P$ checks each block with the following conditions $AdrK(\nu,j)=t-1$ and $AdrW(\nu,j)=i$.
 If this condition is true then $P$ computes   $Val(\nu,i,t-1)$ by this $j$-th block. The result of computation by this $j$-th block
  is the value of $Step_1(\nu,t-1)$.  If this condition is false $P$ goes to next block without branching.

Note that computing of $Val(\nu,i,t-1)$ does not depend on $i$ if we know $j$. And it means the part for computing of $Val(\nu,i,t-1)$ is common for different $i$.

\begin{center}
\begin{figure}[tbh]
    \includegraphics{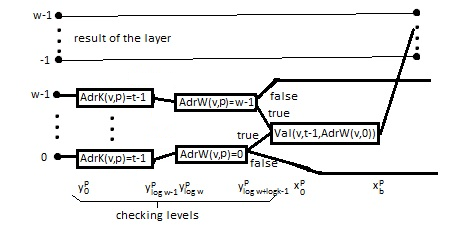}
  \caption{$p$-th block of layer $2t-1$}
\end{figure}
\end{center}

In each level program $P$ has $w+1$ nodes for result of layer. After
computing of $Step_1(\nu,t-1)$ by block $j$ program $P$ goes to one
of result of layer nodes. From result of layer nodes $P$ goes to end
of layer without branching, because result of layer is already
obtained. If block $j$ such that $AdrK(\nu,j)=t-1$ and
$AdrW(\nu,j)=i$ are not founded then $P$ goes to $-1$ result of
layer node and from this node $P$ goes to $0$ result of program node
without branching.

Let us consider layer $2t$. The first level has $w$ nodes for store
each value of function $Step_1(\nu,t-1)$. For $i$-th  node of first
level program $P$ checks each  block for the following condition
$AdrK(\nu,j)=t-1$ and $AdrW(\nu,j)=i+w$. If this condition is true
then $P$ computes   $Val(\nu,i+w,t-1)$ by this $j$-th block. The
result of computation by this $j$-th block is the value of
$Step_2(\nu,t-1)$.  If this condition is false $P$ goes to next
block without branching.

In each level program $P$ has $w+1$ nodes for result of the layer.
After computing of $Step_2(\nu,t-1)$ by block $j$ program $P$ goes
to one of result of layer nodes.

In last layer program $P$ computes  $Val(\nu,i+w,k-1)$ and if
$Val(\nu,i+w,k-1)=0$ then $P$ answers $0$ and answers $1$ otherwise.

Let us compute width of program. The block checking procedure needs
only $2$ nodes in level. Hence for each value of $i$ we need $2w$
nodes in checking levels. Computing of  $Val(\nu,i,t-1)$ and
$Val(\nu,i+w,t-1)$ needs $w$ nodes in non checking levels. And $w$
nodes for going to next block in case the block is not needed for
non checking levels. And result of layer nodes needs $w+1$ nodes.
Therefore we have at most $3w+1$ nodes on each layer.
\Endproof

From paper \cite{ak13} we have the following lower bound.

\begin{theorem}[\cite{ak13}]\label{th-main-ak13}
Let function $f(X)$ is computed  by  $k$-OBDD $P$ of  width $w$,
then $N(f) \leq w^{(k-1)w+1}. $
\end{theorem}

Finally we complite  the proof of Theorem \ref{h-kobdd}. It is
obvious that $\mathsf{k-OBDD_{\lfloor
w/16\rfloor-3}}\subseteq\mathsf{k-OBDD_{w}}$. Let us show inequality
of this classes. Let us look at function $SAF_{\lceil
k/3\rceil,\lceil w/4 \rceil}$. By Lemma \ref{fkw2} we have
$SAF_{\lceil k/3\rceil,\lceil w/4 \rceil}\in \mathsf{k-OBDD_{w}}$.
By Lemma \ref{fkw1}  $N(SAF_{\lceil k/3\rceil,\lceil w/4
\rceil})\geq(\lceil w/4 \rceil)^{(\lceil k/3\rceil-1)(\lceil w/4
\rceil-2)}$.

Let us compute $N(SAF_{\lceil k/4\rceil,\lceil w/5 \rceil}) /(\lfloor w/16\rfloor-3)^{(k-1)(\lfloor w/16\rfloor-3)+1}$.

\[\frac{N(SAF_{\lceil k/3\rceil,\lceil w/4 \rceil})}{(\lfloor w/16\rfloor-3)^{(k-1)(\lfloor w/20\rfloor-3)+1}}
\geq\frac{(\lceil w/4 \rceil)^{(\lceil k/3\rceil-1)(\lceil w/4 \rceil-2)}}{(\lfloor w/16\rfloor-3)^{(k-1)(\lfloor w/16\rfloor-3)+1}}=\]
\[=2^{(\lceil k/3\rceil-1)(\lceil w/4 \rceil-2)\log (\lceil w/4 \rceil) - ((k-1)(\lfloor w/16\rfloor-3)+1)\log (\lfloor w/16\rfloor-3)}\geq\]
\[
\geq 2^{(\lceil k/3\rceil-1)(\lceil w/4 \rceil-2)\log (\lceil w/4 \rceil) - (k-1)(\lfloor w/16\rfloor-2)\log (\lfloor w/16\rfloor-3)}>\]
\[>2^{\frac{1}{4}(k-1)(\lceil w/4 \rceil-2)\log (\lceil w/4 \rceil) - (k-1)(\lfloor w/16\rfloor-2)\log (\lfloor w/16\rfloor-3)}>\]
\[>2^{(k-1)(\lceil w/16 \rceil-2)\log (\lceil w/4 \rceil) - (k-1)(\lfloor w/16\rfloor-2)\log (\lfloor w/16\rfloor-3)}>1\]

Hence $N(SAF_{\lceil k/3\rceil,\lceil w/4 \rceil})>(\lfloor
w/16\rfloor-3)^{(k-1)(\lfloor w/16\rfloor-3)+1}$ and  by Theorem
\ref{th-main-ak13} we have $SAF_{\lceil k/3\rceil,\lceil w/4
\rceil}\notin \mathsf{k-OBDD_{\lfloor w/16\rfloor-3}}$. \Endproof

\end{document}